\documentclass[a4paper,10pt,twoside]{cpc-hepnp}

\usepackage{multicol}
\usepackage{graphicx}
\usepackage{booktabs}
\usepackage{amssymb,bm,mathrsfs,bbm,amscd}
\usepackage[tbtags]{amsmath}
\usepackage{lastpage}
\usepackage{CJKutf8}
\usepackage[colorlinks,linkcolor=blue,anchorcolor=blue,citecolor=blue,urlcolor=blue,filecolor=black]{hyperref}

\begin{document}
\begin{CJK}{UTF8}{gbsn} 



\title{Spin determination from the in-plane angular correlation analysis for \\
various coordinate systems
}

\author{%
      Biao Yang(杨彪)$^{1}$
\quad Yan-Lin Ye(叶沿林)$^{1;1)}$\email{yeyl@pku.edu.cn}
\quad Jian-Ling Lou(楼建玲)$^{1}$
\quad Xiao-Fei Yang(杨晓菲)$^{1}$\\
\quad Jing-Jing Li(李晶晶)$^{1}$
\quad Yang Liu(刘洋)$^{1}$
\quad Wei Liu(刘威)$^{1}$
\quad HanZhou Yu(余翰舟)$^{1}$
}
\maketitle

\address{%
$^1$ School of Physics and State Key Laboratory of Nuclear Physics and Technology, Peking University, Beijing 100871, China \\
}

\begin{abstract}
In a reaction to excite the resonant state followed by the sequential cluster-decay, the in-plane angular correlation method is usually applied to determine the spin of the mother nucleus. However, the correlation pattern exhibited in a two-dimensional angular-correlation spectrum depends on the selected coordinate system. Particularly the parity-symmetric and the axial-symmetric processes should be presented in a way to enhance the correlation pattern whereas the non-symmetric process should be plotted elsewhere in order to reduce the correlation background. In this article, three possible coordinate systems, which were previously adopted in the literature, are described and compared to each other. The consistency of these systems is evaluated based on the real experimental data analysis for the 10.29-MeV state in $^{18}$O. A spin-parity of 4$^+$ is obtained for all three coordinate systems.
\end{abstract}



\begin{multicols}{2}
\section{Introduction}

Nuclear clustering is currently one of the hot topics in nuclear structure studies, especially for nuclei in the expanded nuclear chart away from the $\beta$-stability line \cite{Yang2014,Lyu2016,SCYang2014,Jiang2017,Li2017,Feng2019,Zang2018}. The cluster structure is often formed and stabilized at highly excited resonant states at the vicinity of the cluster-separation threshold \cite{Ikeda1968PTPS}. Experimentally this structure can be probed via nuclear reaction tools combined with sequential cluster-decay. The reconstruction of the resonant state from the decay fragments, the so-called invariant mass (IM) method, allows to select the states with large cluster partial decay widths. This selection reduces significantly the level density at high excitation energies, being in favor of the quantitative extraction of the physical properties of the resonances. Furthermore, a model independent determination of the spin of the reconstructed resonant state can be achieved though the sensitive angular correlation method \cite{RAE1984_1,Freer1996}. This is particularly important at high excitation energies where the differential cross section method becomes almost useless due to the overlap of many close-by states and the uncertainties in many fitting parameters~\cite{RAE1984_1}.

Determination of the spin of each resonant state is of particular importance in order to form the molecular band which is required to firmly establish the clustering structure \cite{Oertzen2006PR}. So far the angular correlation analysis has been the most sensitive and reliable tool to determine the spin of the resonant state \cite{Pougheon1979,Freer1996,Chappell1996,Curtis2001,Charity2015}. Nevertheless, the angular correlation picture depends on the selection of the coordinate system, which varies for different types of experiments and conventions of data analysis. Basically, although in most cases, the z-coordinate axis is fixed on the beam direction, the definition of the spherical angle axis differs from each other in these coordinate systems. So far no detailed analysis and comparison of these coordinate systems were provided in the literature, which may lead to some misunderstanding and erroneous application of the angular correlation method.

In this work we systemically investigate three kinds of coordinate systems which have been usually used in the literature. The definitions and features of these systems are outlined and compared to each other. The consistency of these systems is demonstrated by the real experimental data analysis for the \mbox{10.29-MeV} resonant state in $^{18}$O. Some suggestions for the application of the angular correlation method are provided in the summary session.

\section{In-plane angular correlation}
\subsection{General description}

For a sequential decay reaction a(A,B$^*\to$ c + C)b, the composite resonant particle B$^*$ may decay into, for instance, two spin-zero fragments. The angular correlation of the latter is a sensitive probe of the spin of the resonant state in the mother nucleus B \cite{Freer1996}. In a spherical coordinate system with its z-axis pointing to the beam direction (Fig.\ref{angle}), the correlation function can be parameterised in terms of four angles \cite{Freer1996}. Namely, the center-of-mass (c.m.) scattering polar and azimuthal angles, $\theta^*$ and $\phi^*$, respectively, of the resonant particle B$^*$; the polar and azimuthal angles, $\psi$ and $\chi$, respectively, of the relative velocity vector $v_{\textrm{rel}}$ of the two fragments (the arrow connecting HI and LI in Fig.~\ref{angle}). Both polar angles, $\theta^*$ and $\psi$, are with respective to the beam direction. The azimuthal angle $\phi$ takes 0$^\circ$ (or 180$^\circ$) in the horizontal plane defined by the center positions of the detectors placed at the opposite sides of the beam (the chamber plane or the detection plane). Another azimuthal angle $\chi$ is defined to be 0$^\circ$ (or 180$^\circ$) in the reaction plane fixed by the beam axis and the reaction product B$^*$. Owing to the limited detector geometry in a general experiment, the correlation is often approximately constrained in the chamber plane as shown in Fig.~\ref{angle}. In this case, the azimuthal angles $\phi$ and $\chi$ remain always at 0$^\circ$ or 180$^\circ$, depending on the selected coordinate system, and the angular correlation appears as a function of only two polar angles $\theta^*$ and $\psi$. This is called the in-plane correlation.

\begin{center}
\includegraphics[width=0.4\textwidth]{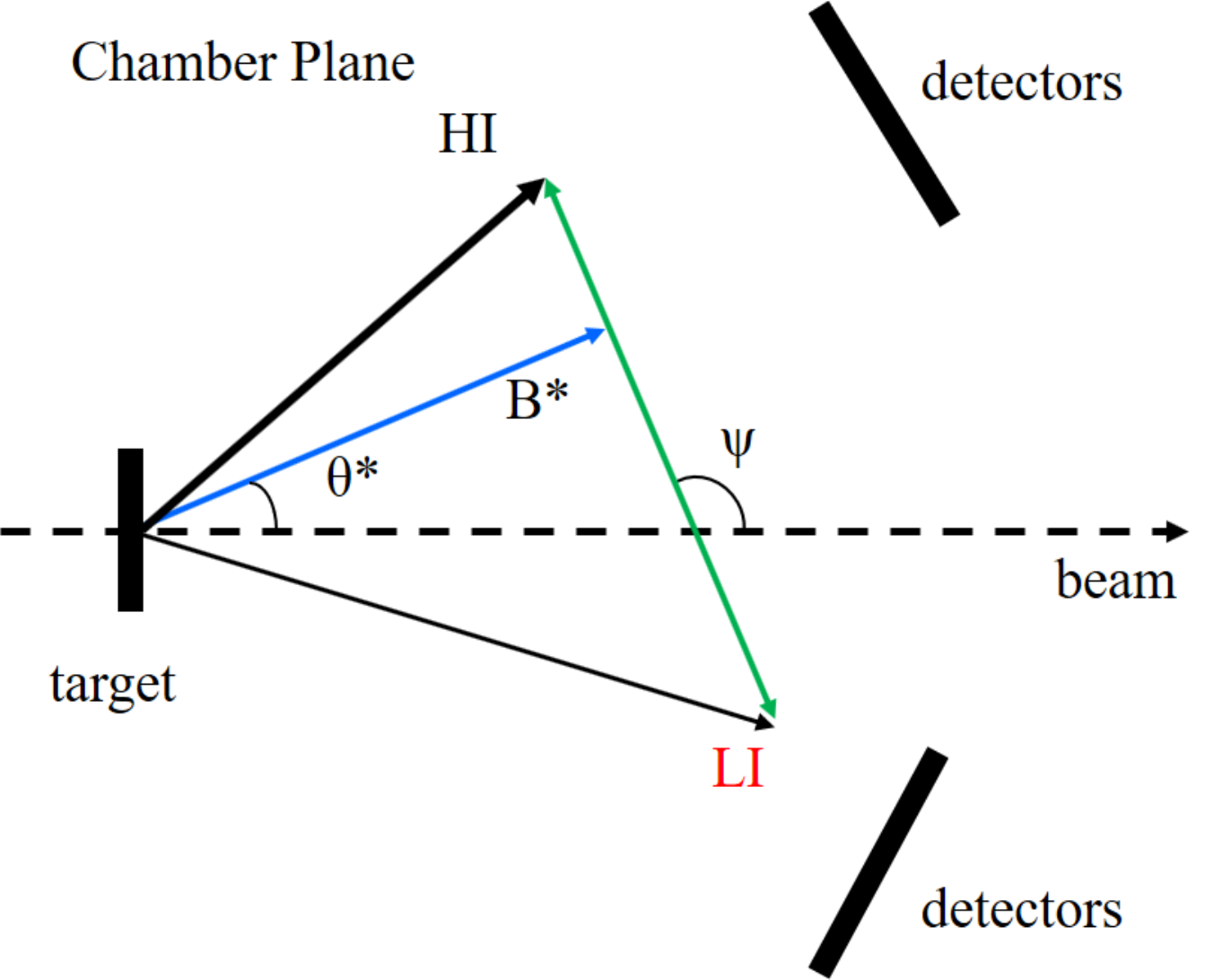}
\figcaption{The schematic diagram of the sequential decay following a transfer reaction a(A,B$^*\to$ c + C)b and the definitions of the polar angles $\theta^*$ and $\psi$ in the detection plane. The two decay fragments, c and C, are specified as light ion (LI) and heavy ion (HI), respectively, in the figure.}
\label{angle}
\end{center}

When the azimuthal angle $\chi$ is restricted to 0$^\circ$ (or 180$^\circ$ ), the most striking feature of the angular correlation in $\theta^*$-$\psi$ plane appears as the ridge structures associated with the spin of the mother nucleus (\cite{RAE1984_1,Marsh1985,Freer1996}). At relatively small $\theta^*$ angles, the structure is characterized by the locus $\psi = \alpha\theta^*$ in the double differential cross sections, where $\alpha$ is a constant for the slope of the ridge and nearly inversely proportional to the spin of the resonant state B$^*$ \cite{Freer1996,Marsh1985}. The correlation is oscillatory along $\psi$ angle for a fixed $\theta^*$, and vice versa. In general, this in-plane correlation structure can be projected onto the one-dimensional spectrum \mbox{ $W(\theta^*=0^{\circ},\psi'=\psi - \alpha\theta^*)$}. Within the strong absorption model (SAM) \cite{Bond1980,Frahn1980,Frahn1980NPA}, $\alpha$ may be related to the orbital angular moment $l_{i}$ of the dominant partial wave in the entrance channel, through $\alpha=\frac{l_{i}-J}{J}$, with $J$ being the spin of B$^*$ \cite{Marsh1985,RAE1984_1}. $l_{i}$ can be evaluated simply from $l_{i} = r_{0}(A^{1/3}_{p} + A^{1/3}_{t})\sqrt{2{\mu}E_{\textrm{c.m.}}}$~\cite{Curtis2002}, with $A_{p}$ and $A_{t}$ the mass numbers of the beam and target nucleus, respectively, $\mu$ the reduced mass and $E_{\textrm{c.m.}}$ the center-of-mass energy. If the resonant nucleus is emitted to angles close to \mbox{$\theta^{*}$ = 0$^{\circ}$}, the projected correlation function \mbox{ $W(\psi')$} is simply proportional to the square of the Legendre polynomial of order $J$, namely $|P_{J}(\rm{cos}(\psi))|^{2}$. This method has been frequently applied in the literature (\cite{Aquila2016} and references therein) and will also be demonstrated in the following section 2.3.

\subsection{Different coordinate systems}

As indicated above, what's important in the application of the angular correlation plot is to enhance as clear as possible the ridge structure which corresponds to the spin of the resonant mother nucleus. In this sense the selection of the coordinate system for the plot is meaningful. For an unpolarized experiment, the reaction process satisfies the axial symmetry around the beam axis. In addition the decay process should satisfy the parity (space inversion) symmetry with respective to the c.m. of the resonant mother nucleus. In the two-dimensional correlation plot (plot of the double differential cross section) as functions of the polar angles $\theta^*$ and $\psi$, it is natural to plot these symmetrical events in the same ridge band while placing the non symmetric events elsewhere, in order to enhance the sensitivity to the associated spin. This care should be taken even for the simple case of restricted detection around the chamber plane. For instance, in Fig.~\ref{different} we show schematically two processes which generate the same polar angles $\theta^*$ and $\psi$ but are not symmetric in terms of the resonance-decay. These two processes should be differentiated from the ridge plot by using an appropriate coordinate system, such as the one using positive and negative $\theta^*$ as defined below. On the other hand, the coordinate definition should keep the four symmetric processes, as schematically displayed in Fig.~\ref{equivalent}, in the same ridge band so that the ridge structure appears continuously and a simple projection may bring them together to enhance the sensitivity to the associated spin. We introduce in the following three kinds of coordinate systems and show their differences in plot-definition and the consistency in extraction of the spin.

\begin{center}
\includegraphics[width=0.4\textwidth]{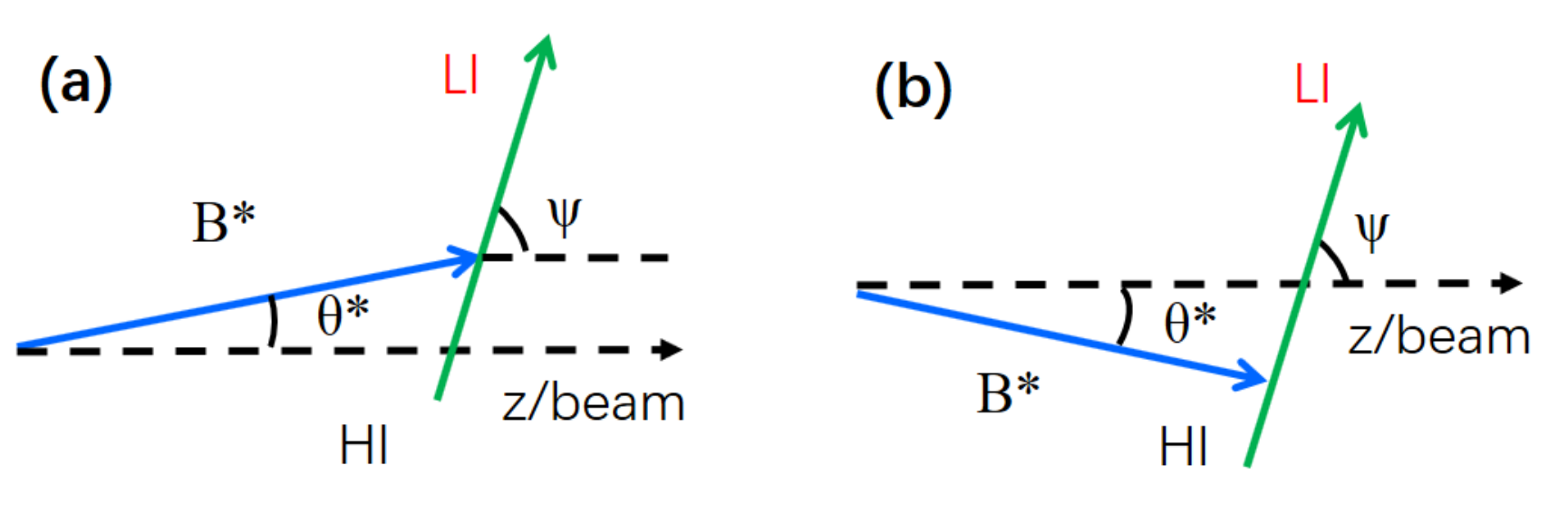}
\figcaption{The schematic diagram of the two independent reaction processes having the same polar angles. The direction of the relative velocity vector $v_{\textrm{rel}}$ is guided by the light ion for both cases.}
\label{different}
\end{center}

\end{multicols}
\begin{center}
\includegraphics[width=0.8\textwidth]{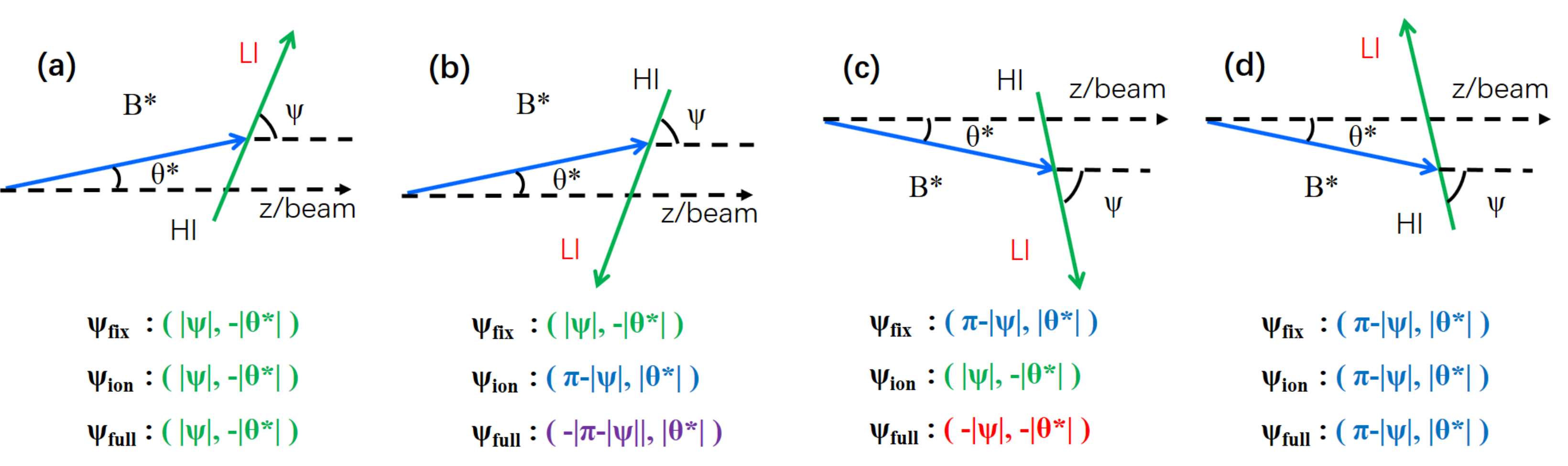}
\figcaption{Schematic diagram of the four symmetric reaction-decay processes in the chamber plane. (a) and (b) are parity-symmetric processes while (c) and (d) are their axial-symmetric processes, respectively. All processes are identified by the angles $\theta^*$ and $\psi$ defined in various coordinate systems as described in the text.}
\label{equivalent}
\end{center}

\begin{center}
\includegraphics[width=0.8\textwidth]{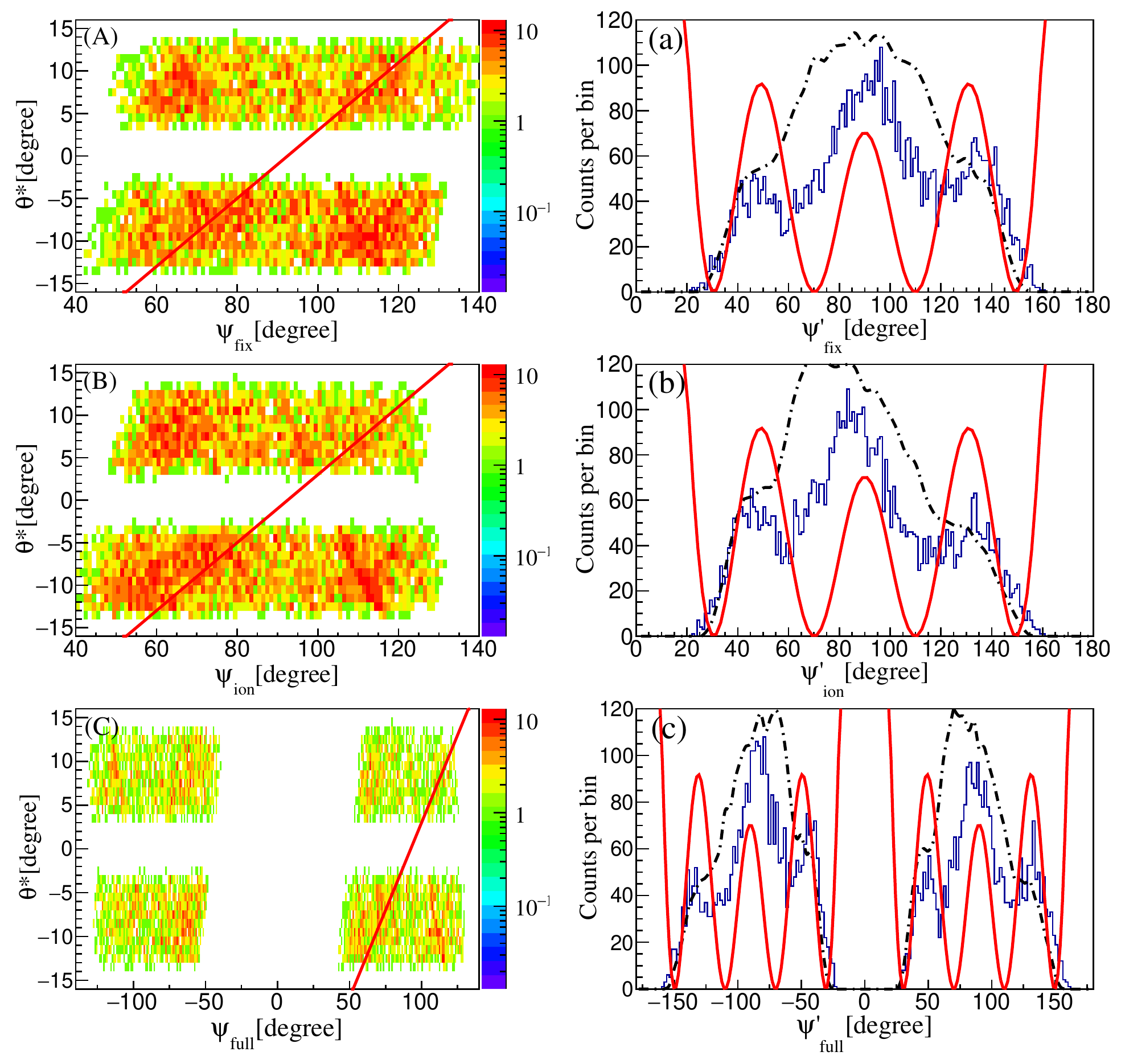}
\figcaption{(Color online) Angular correlation spectra for the 10.29-MeV state in $^{18}$O using the (A) $\psi_{\textrm{fix}}$, (B) $\psi_{\textrm{ion}}$ and (C) $\psi_{\textrm{full}}$ coordinate systems, respectively. The projection lines in (A), (B) and (C) (the red-solid lines) correspond to a slope parameter $\alpha~=~\frac{l_{i}-J}{J}$, with $l_{i}$~=~13.9 $\hbar$ (using $r_{0}$~=~1.2 fm). (a), (b) and (c) show the projections of (A), (B) and (C) onto $\theta^*$~=~0$^\circ$ axis, respectively. The black dot-dashed line indicates the simulated detector efficiency in each coordinate system. All experimental distributions are compared with a squared Legendre polynomial of order 4.}
\label{correlation}
\end{center}
\begin{multicols}{2}

In the first coordinate system (hereinafter named $\psi_{\textrm{fix}}$ coordinate system), the relative velocity vector of the decay products always points to the fixed detector at one side of the beam (where $\phi^*$ = $0^\circ$). This is practical when two decay particles are identical, such as $^{24}\rm{Mg}~\rightarrow~^{12}\rm{C}+^{12}\rm{C}$ \cite{Freer1996}, or each detector is designed to be sensitive to only one type of the particles, such as $^{18}\rm{O}~\rightarrow~^{14}\rm{C}+\alpha$ with $^{14}\rm{C}$ always detected at one-side while $\alpha$ at another side, of the beam \cite{RAE1984_1}. By definition $\theta^*$ is positive on the opposite side of the beam and negative on the same side, in comparison to the fixed positive $\psi$. With this definition, processes (a) and (b) will be plotted at one position (negative $\theta^*$) while (c) and (d) at another position (positive $\theta^*$) (see present Fig.~\ref{correlation}(A) or Fig.~5 in Ref.~\cite{Freer1996}).

In another more ``physical'' convention \cite{Marsh1985} (hereinafter denoted as $\psi_{\textrm{ion}}$ coordinate system), the relative velocity vector $v_{\textrm{rel}}$ is always pointing to a certain decay particle (usually the lighter one, LI), corresponding to $\phi^*$ = $0^\circ$ . The positive $\theta^*$ remains at the opposite side of the positive $\psi$. Under this convention the axial-symmetric processes (a) and (c) will be plotted at the same position, whereas (b) and (d) at another position, as demonstrated in the present Fig.~\ref{correlation}(B) or in Fig.~3 of Ref.~\cite{Marsh1985}. Due to the different detection efficiency for light and heavy particles, the correlation structure in this coordinate system may differ from that in the $\psi_{\textrm{fix}}$ coordinate system.

Additionally, on the basis of the $\psi_{\textrm{ion}}$ coordinate system, the polar angle $\psi$ could also be assigned positive or negative sign, depending on the azimuthal angle $\chi$. Firstly $0^\circ$ of the azimuthal angle $\phi^*$ or $\chi$ is defined by one detector in the chamber plane. Then positive $\psi$ means $\chi$ close to $0^\circ$ while negative $\psi$ for $\chi$ close to $180^\circ$. Now four intrinsically equivalent cases in Fig.~\ref{equivalent} will be plotted at four different positions in the $\theta^*$-$\psi$ plane. This coordinate system is denoted as $\psi_{\textrm{full}}$. This convention was used in some literature, such as in Ref.~\cite{Curtis2001}, and also demonstrated in the present Fig.~\ref{correlation}(C). Now we have both $\theta^*$ and $\psi$ ranging in positive as well as in negative scales. Since the experimental detection system may not be exactly symmetric with respective to the beam axis, the double differential cross section in Fig.~\ref{correlation}(C) seems not symmetric neither. It is evident that this wider scale distribution would give more consistent information for the ridge structure, but in the mean time would require higher statistics.

We note that the above introduced three coordinate systems are equally meaningful since the non-symmetric process, as shown in Fig.~\ref{different}(b), would appear in neither of the defined ridge band. In the mean time these systems should be consistent to each other in terms of extracting the spin of the resonant mother nucleus. This consistency is demonstrated below by real experimental data analysis.

\subsection{Experimental results}
Recently, a multi-nucleon transfer and cluster-decay experiment $^{9}$Be($^{13}$C,$^{18}$O$^{*}~\to~^{14}$C~+~$\alpha$)$\alpha$ was performed at the HI-13 tandem accelerator facility at China Institute of Atomic Energy (CIAE) in Beijing. Resonant states in $^{18}$O can be reconstructed according to the invariant mass method \cite{Yang2014,Li2017}, as shown in Fig.~\ref{excitation} for events with small $\theta^*$ angles. The state at 10.29 MeV is a good candidate for angular correlation analysis, owing to its clear peak identification and relatively large $\psi$-angle coverage. In Fig.~\ref{correlation} we plot the angular correlation spectrum for the $^{18}$O 10.29-MeV state in the above described three coordinate systems (Fig.~\ref{correlation}(A-C)). Further these two dimensional spectra in $\theta^*-\psi$ plane are projected onto the $\theta^*$~=~0$^\circ$ axis according to the above described \mbox{ $\psi'=\psi - \frac{l_{i}-J}{J}\theta^*$} relation, as exhibited in Fig.~\ref{correlation}(a-c), respectively. The projections are compared with the square of the Legendre polynomial of order 4. It would be worth noting that only the periodicity of the distribution matters whereas the absolute peak amplitudes depends on the detection efficiency. Although the distributions behave slightly different among three coordinate systems, the periodicities of the experimental spectra all agree with the Legendre polynomial of order 4, corresponding to a spin-parity of 4$^+$ for the 10.29-MeV state in $^{18}$O. This consistency between various coordinate systems indicates the reliability of the angular correlation method in determining the spin of a resonant state.

\begin{center}
\includegraphics[width=0.4\textwidth]{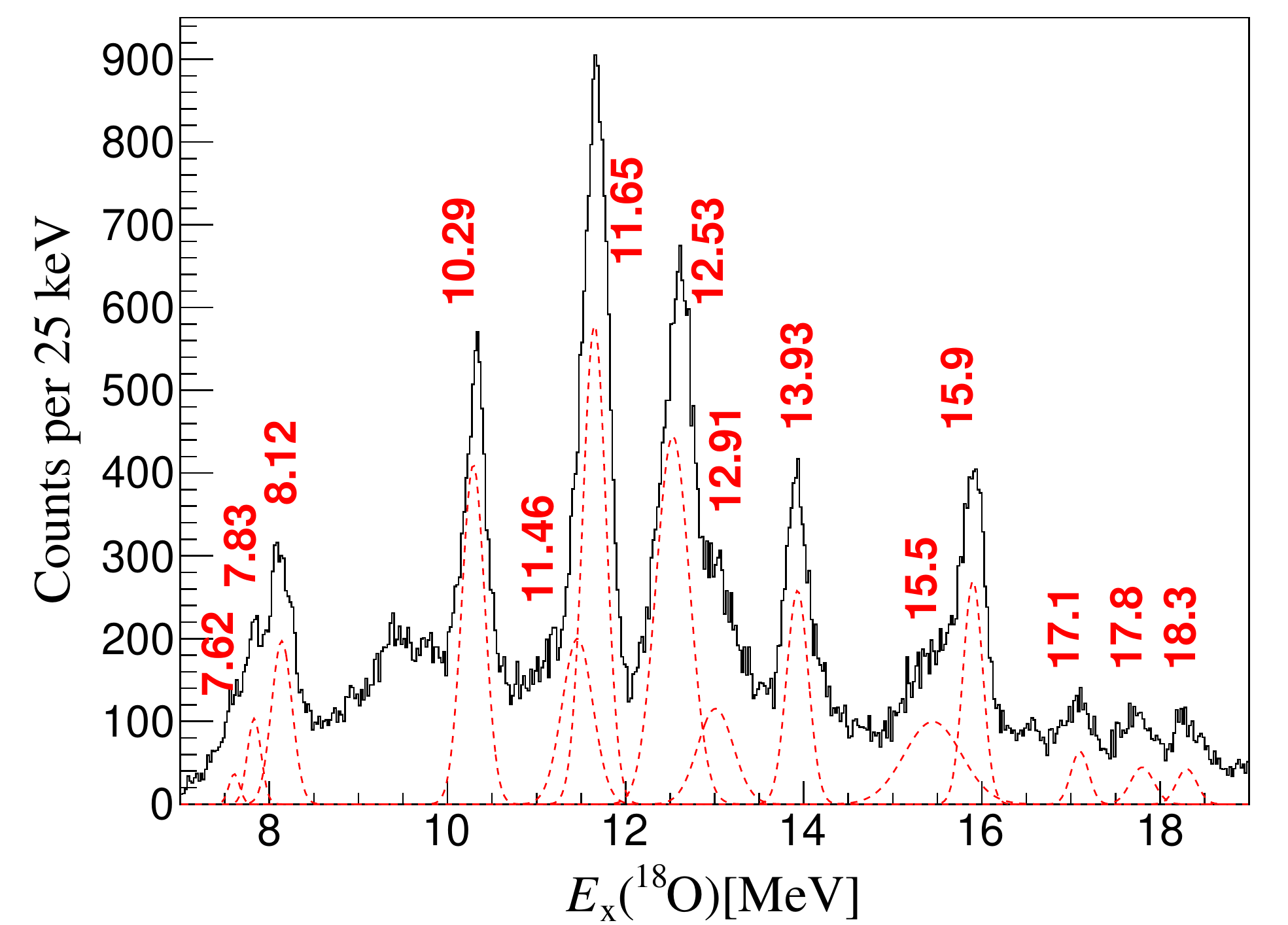}
\figcaption{Excitation energy spectrum for $^{18}$O emitting to small $\theta^*$ angles. The state at 10.29 MeV is selected to demonstrate the angular correlation analysis.}
\label{excitation}
\end{center}

Based on the consistency exhibited in Fig.~\ref{correlation} and the symmetry property of the Legendre polynomial, we may plot the projected correlation spectrum as a function of $|\textrm{cos}(\psi')|$ \cite{Aquila2016}, in order to increase the statistics in each bin of the distribution. Furthermore, this plot is independent of the above defined coordinate systems. In addition we may reconstruct the excitation energy spectrum, similar to that in Fig.~\ref{excitation}, for each bin of $|\textrm{cos}(\psi')|$ and extract the corresponding event number for the pure 10.29-MeV peak by subtracting the smooth background beneath the peak. The experimental correlation spectrum is now plotted in Fig.~\ref{spin103}. The theoretical function composed of a squared Legendre polynomial plus a constant background, corrected by the detection efficiency, is used to describe the experimental results. Now, we can see that, not only periodicity, but also the magnitude of the function for spin-parity of $4^+$ give rise to an excellent fit to the experimental data, whereas other options of spin-parity can be excluded. We notice that a constant background is still needed in the theoretical function since the experimental data include some uncorrelated components coming from events within the \mbox{10.29-MeV} peak but away from the exact $\theta^*$~=~$0^\circ$ axis \cite{Freer1996}.

\begin{center}
\includegraphics[width=0.4\textwidth]{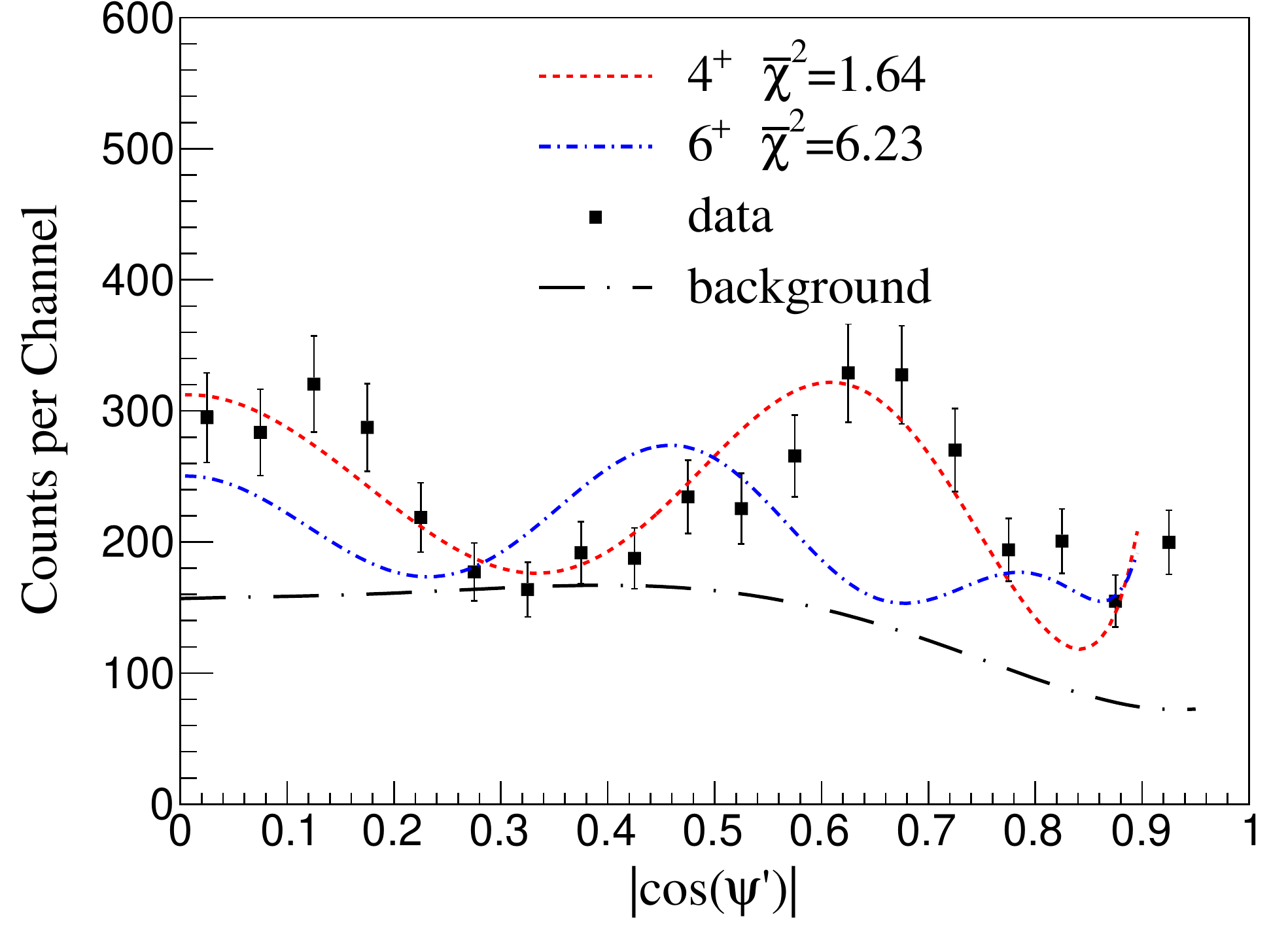}
\figcaption{The angular correlation spectrum for 10.29-MeV state in $^{18}$O, in comparison to the Legendre polynomials of order 4 (the red dotted line) and order 6 (the blue dot-dashed line). A uniformly distributed background is assumed to account for the uncorrelated component (the long dot-dashed line).  Theoretical angular distributions has been corrected for the detection efficiency. The corresponding reduced $\bar{\chi}^2$ for the goodness of the theoretical description is also indicated in the plot.}
\label{spin103}
\end{center}

\section{Summary}
The fragment angular correlation in a sequential cluster-decay reaction provides a model-independent way to determine the spins of the resonant nucleus. When the correlation spectrum is restricted to angles close to the detection chamber plane, the ridged pattern can be clearly exhibited in the two dimensional plot with respective to the two polar angles $\theta^*$ and $\psi$. According to the ways to deal with the symmetrical events, three coordinate systems for different $\theta^*$ and $\psi$ definitions have been adopted in the literature for various experiments and spin analysis. In the present work, we outlined these systems and compared them to each other to clarify their differences and consistency. These systems are examined by the cluster-decay data for 10.29-MeV state in $^{18}$O, measured in our recent experiment. This work gives a better understanding of the angular correlation function, and demonstrates the possible choices for the best extraction of the spin of a resonant state.

In the case that resonant nucleus decays into two spin-zero fragments, the two-dimensional correlation spectrum for small $\theta^*$ angles can be projected onto the $\theta^*$~=~$0^\circ$ axis. This projected spectrum may be described by a squared Legendre polynomial with an order corresponding to the spin of the resonant mother nucleus. Using this method, a spin-parity of 4$^+$ is firmly determined for 10.29-MeV state in $^{18}$O.

Based on the above investigations, we might propose the basic procedure for the application of the angular correlation method. Firstly, the experiment should be designed to have good detection for events at small $\theta^*$ angles and with wide $\psi$ coverage. Secondly, a proper coordinate system should be selected according to the detection and data-distribution characteristics. In principle, the best choice is the $\psi_{\textrm{full}}$ coordinate system owing to its wider angular range of the correlation spectrum which may help to identify the ridge structure as long as the statistics is good enough.  However, when the statistics is low, the coordinate systems $\psi_{\textrm{fix}}$ or $\psi_{\textrm{ion}}$ , depending on the detection arrangement, is more convenient. Thirdly, before the projection onto the $\theta^* = 0^\circ$ axis, the ridge structure should have been distinctly observed. Otherwise, any small shift in projection direction may lead to wrong extraction of the spin, especially in the case of higher spin where more oscillations in the projected spectrum are expected. The detector efficiency should also be carefully examined, since the large variation of the efficiency curve may give rise to some non-physical structure in the projected distribution. Finally, the projected spectrum may be obtained by subtracting the background for each bin of $|\textrm{cos}(\psi')|$. This real experimental spectrum can be compared with the theoretical function (Legendre polynomial corrected by the detection efficiency) and the quantitative goodness analysis can be examined. But again the correct projection parameter should be fixed before this fitting procedure.

\vspace{ 5 mm}
\acknowledgments
{This work was supported by the National Key R$\&$D Program of China (Grant No. 2018YFA0404403), and the National Natural Science Foundation of China (Grant Nos. 11535004, 11875074, 11775004, 11875073).
}

\end{multicols}

\vspace{-1mm}
\centerline{\rule{80mm}{0.1pt}}
\vspace{2mm}

\begin{multicols}{2}
\bibliographystyle{apsrev4-1}
\bibliography{ref}
\end{multicols}

\end{CJK}  

\end{document}